\documentclass[prb,twocolumn,floatfix,showpacs]{revtex4}
\usepackage{amsfonts}
\usepackage{amsmath}
\usepackage{bm}
\usepackage{graphicx}
\begin{document}
\title{Angle-resolved photoemission spectra in the cuprates  from  the $d$-density wave theory}
\author{Sudip Chakravarty, Chetan Nayak, and Sumanta Tewari}
\affiliation{Department of Physics and Astronomy, University of California Los Angeles, Los Angeles, CA 90095-1547}
\date{\today }
\begin{abstract}
Angle-resolved photoemission spectra present
two challenges for the $d$-density wave (DDW) theory of the
pseudogap state of the cuprates: (1) hole pockets
near $(\pi/2,\pi/2)$ are not observed, in apparent contradiction with the
assumption of translational symmetry breaking,
and (2) there are no well-defined
quasiparticles at the {\it antinodal}
points, in contradiction with the predictions of
mean-field theory of this broken symmetry  state. Here,
we show how these puzzles can be resolved.
\end{abstract}
\pacs{}
\maketitle

At first glance, the $d$-density wave (DDW) proposal for the pseudogap
state of the cuprates\cite{CLMN} seems to naturally explain
the principal anomaly in photoemission spectra
in this state: the existence of a gap with
$d_{x^2 -y^2}$ symmetry without superconductivity.
However,  since DDW order breaks translational
symmetry, thereby splitting the Brillouin zone into
two magnetic Brillouin zones, the Fermi surface in the
first magnetic Brillouin zone should be duplicated in the second
magnetic Brillouin zone. Thus, the Fermi surface consists of hole pockets, which is of importance in understanding a number of experiments, such as superfluid density\cite{Sumanta1}, Hall number\cite{Sumanta2}, etc. However, in angle-resolved
photoemission spectroscopy (ARPES) in hole doped cuprates,
Fermi arcs -- not hole pockets -- are observed \cite{RMP}.
There is spectral weight in the first, but not
the second magnetic zone. In this paper, we show from a careful analysis that
Fermi arcs rather than hole pockets are indeed the
consequences of the DDW theory in ARPES.  

A second important aspect of the proposal of
a broken symmetry state, even one of an unusual
variety, is that it is expected to support electronic quasiparticles
which are essentially Fermi-liquid like,
as they are in a BCS superconductor. However, from ARPES in underdoped samples
no peak is observed at the antinodal points in the normal state,
but one appears in the superconducting state upon cooling.\cite{RMP}
This observation also finds a natural explanation within 
our theory.\cite{pi0} We show that the antinodal
quasiparticles, being relatively high-energy excitations,
decay by creating particle-hole
pairs along the Fermi arcs in the DDW state. In contrast,  in the $d$-wave 
superconducting state (DSC), or in the coexisting DDW and DSC state, the Fermi arcs shrink
to points, and the decay rate is considerably reduced, resulting in a peak in the spectral function.
This reduction is bolstered by the suppression
of the decay matrix element by the superconducting
coherence factors. 

The explanation discussed here involves interaction between quasiparticles, whose absolute magnitude is set by a reasonable Hubbard-like interaction of magnitude 1.5 eV, but the precise magnitude is of not much consequence.  There may be other sources of broadening of the quasiparticle peak, including fluctuation effects, bilayer splitting, fractionalization, etc., which we do not address here. We merely wish to point out that within the simplest mean field picture of  DDW, there are {\it no} obvious puzzles.

In our mean field analysis, and indeed in many theories,  the nodal quasipartcles, or excitations at the Fermi arcs, should in principle be {\it sharp}, which is not entirely 
in keeping with ARPES, although a fairly well defined peak is observed both above and below the superconducting transition temperature, $T_{c}$. It remains to be seen if the present experimental situation changes with time or not.

To establish our notation, we begin with a brief summary of the mean field theory of DDW. The Hamiltonian ${\cal H} $ can be simply  written  in the first magnetic zone by introducing the Pauli matrices $\sigma_{x}$, $\sigma_{z}$, the identity matrix $\mathbb{I}$, a row vector $\Psi_{\mathbf{k},\alpha}^{\dagger}\equiv(c_{\mathbf{k},\alpha}^{\dagger}, -ic_{\mathbf{k+Q},\alpha}^{\dagger})$, and its Hermitian adjoint. The electron destruction  operators of  momentum $\bf k$ and spin $\alpha$ are $c_{{\mathbf k},\alpha}$ and the momentum ${\bf Q}=(\pi,\pi)$. Thus, ${\mathcal K}={\mathcal H} - \mu {\mathcal N}$ is given by
\begin{equation}
{\mathcal K}=\sum_{\mathbf{k},\alpha}\Psi_{\mathbf{k},\alpha}^{\dagger}\left[\left(\epsilon_{\mathbf{k}}^{+}-\mu\right)\mathbb{I}+\epsilon_{\mathbf{k}}^{-}\sigma_{z}+W_{\mathbf{k}}\sigma_{x}\right]\Psi_{\mathbf{k},\alpha} ,
\end{equation}
Here $\mu$ is the chemical potential and $\mathcal N$ is the number operator.
Note that ${\mathcal K}$ is complex Hermitian, reflecting broken time reversal symmetry.  
In the first magnetic zone, it is convenient to define
$\epsilon_{\mathbf{k}}^{\pm}= \frac{1}{2}\left[\epsilon_{\mathbf{k}} \pm \epsilon_{\mathbf{k+Q}}\right]$,
where $\epsilon_{\mathbf{k}}$ is the electronic band structure. A standard Bogoliubov transformation diagonalizes the Hamiltonian, but since $\mathbb{I}$ commutes with $\sigma_{x}$ and $\sigma_{z}$, $\left(\epsilon_{\mathbf{k}}^{+}-\mu\right)$ can not enter the coherence factors, which are 
\begin{equation}
\begin{array}{c}
u_{\mathbf{k}}^{2}\\
v_{\mathbf{k}}^{2}
\end{array}\bigg\}=\frac{1}{2}\,\left(1\pm\frac{\epsilon_{\mathbf{k}}^{-}}{E_{\mathbf{k}}}\right)
\end{equation}
where
$E_{\mathbf{k}}=\sqrt{(\epsilon^{-}_{\mathbf{k}})^{2}+W_{\mathbf k}^{2}}$.
The coherence factors must trade places
as $\mathbf{k}\to \mathbf{k+Q}$, which is a consistency check as to  why they cannot be functions of $\left(\epsilon_{\mathbf{k}}^{+}-\mu\right)$.
The energy eigenvalues are:
\begin{equation}
E_{\bf k}^{\pm}= \epsilon_{\bf k}^{+}
\pm E(\mathbf{k}).
\end{equation}
The DDW gap is assumed to take the form:
\begin{eqnarray}
W_{\mathbf k} = \frac{W_0(T)}{2}\left(\cos{k_x}-\cos{k_y}\right)
\end{eqnarray}

The electron spectral function in a crystal need not be
invariant under translation by a reciprocal lattice
vector. In fact, it is weighted by the Fourier transform of
the relevant Wannier orbitals. If the Wannier orbitals are $\delta$-functions,
the spectral weight is the same in all Brillouin zones.
On the other hand, if the Wannier orbital is spread
out spatially, then the spectral weight in higher
Brillouin zones will be very small, and
$I(\omega,\mathbf{k}+\mathbf{G})\ll I(\epsilon,\mathbf{k})$,
where $I(\epsilon,\mathbf{k})$ is the angle-resolved photoemission
intensity.
In the DDW state, the unit cell has been doubled.
The coherence factors $u_{\mathbf{k}},v_{\mathbf{k}}$ tell us
how the two sites within the unit cell are
superposed, so ${v_{\mathbf k}}/{u_{\mathbf k}}$ plays the role of
the Wannier function. The corresponding spectral function in the DDW state
is:
\begin{equation}
\frac{A(\omega,{\bf k})}{2\pi} = 
u_{\mathbf k}^2\,\delta\left(\omega-E_{{\bf k}}^{+}+\mu\right)
+ v_{\mathbf k}^2\,\delta\left(\omega-E_{\bf k}^{-}+\mu\right)
\end{equation}
Consider $\mu<0$, the case of hole doping, such 
that the chemical potential lies entirely
in the valence band, so that $E_{\bf k}^{+}-\mu>0$.
Then the ARPES intensity is
\begin{equation}
I(\omega,{\bf k}) \propto {n_F}(\omega)
\: v_{\mathbf k}^2\,\delta\left(\omega-E_{\bf k}^{-}+\mu\right)
\end{equation}
Since $v_{{\mathbf k}+\mathbf{Q}}=u_{\mathbf k}$, the photoemission intensity
in the first and second magnetic zones differ
only by these coherence factors:
\begin{eqnarray}
I(\omega,{\bf k}) &\propto& {n_F}(\omega)
\: v_{\mathbf k}^2\,\delta\left(\omega-E_{\bf k}^{-}+\mu\right),\\
I(\omega,{\bf k}+{\bf Q}) &\propto& {n_F}(\omega)
\: u_{\mathbf k}^2\,\delta\left(\omega-E_{\bf k}^{-}+\mu\right).
\end{eqnarray}
For ${\bf k}$ in the first magnetic zone
(i.e. for ${\bf k}+{\bf Q}$ in the second magnetic
zone), $u_{\mathbf k}$ vanishes when $W_{\mathbf k}$ vanishes.
In other words, the photoemission intensity
in the second magnetic zone vanishes along
the diagonals. For wavevectors close to
the diagonals, the intensity goes as
$W_{\mathbf k}^2$. Thus, the outer section
of the hole pockets will have small
or even vanishing spectral weight,
and may not be detected in ARPES
experiments. The spectral weight at a typical
point on the outer part of a
hole pocket will depend on
various details, including the
band structure, the precise
angular dependence of the DDW
gap, etc.

\begin{figure}[htb]
\includegraphics[scale=0.7]{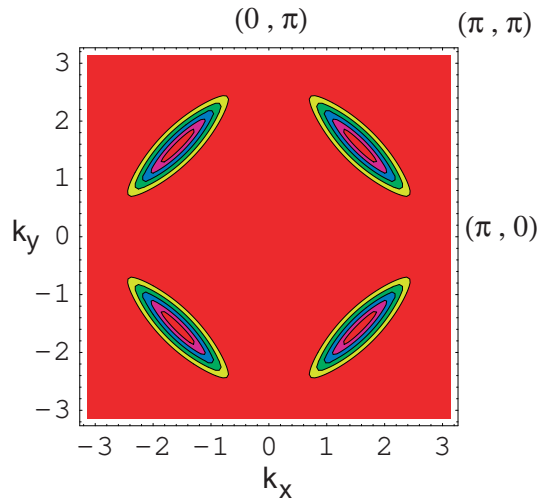}
\caption{The Fermi surface for $W_0(0) = 0.06 \; \text{eV}$. The band structure parameters are defined in the text. }
\label{fig:fs}
\end{figure}

\begin{figure}[htb]
\includegraphics[scale=0.7]{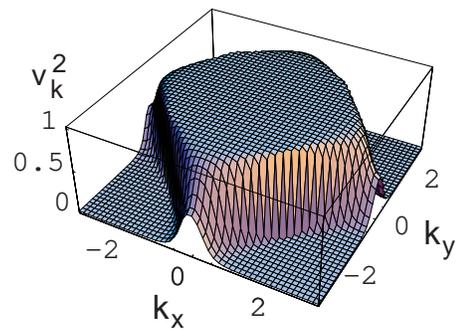}
\caption{The coherence factor $v_{\mathbf k}^2$. The parameters are the same as in Fig.~\ref{fig:fs}.}
\label{fig:vksq}
\end{figure}

A commonly-used model for the band structure is given by
\begin{eqnarray}
\epsilon_{\bf k}^{+}&=&4 t' \cos k_x \cos k_y, \\
\epsilon_{\bf k}^{-}&=&-2 t (\cos k_x +\cos k_y).
\end{eqnarray}
A generic parameter set is  
$t=0.3\; \text{eV}$, $t'/t =0.3$,  $\mu = -0.3\; \text{eV}$; with this set of parameters, the doping level is 14.3\%  The Fermi surface with a typical value of $W_0(0) = 0.06 \; \text{eV}$
consists of four hole pockets as shown in Fig.~\ref{fig:fs}.
The corresponding $v_{\mathbf k}^2$ appearing in the photoemission intensity is shown Fig.~\ref{fig:vksq}.
It is clear that only half of the hole pockets will be visible in the ARPES spectra, resulting in Fermi arcs, despite the fact that the actual Fermi surface consists of hole pockets.\cite{Hamada} The results are similar for the second set of band structure parameters.

We now turn to a discussion of the lifetime of a quasiparticle at the antinodal region ${\bf k}^{*}$ close to $(\pi,0)$, where the free electron Fermi surface crosses the band edge. The equation that determines ${\bf k}^{*}$  is obtained by solving 
\begin{equation}
\epsilon_{{\bf k}^{*}}\equiv\epsilon^{+}_{\pi,k_{y}}+\epsilon^{-}_{\pi,k_{y}}=\mu
\end{equation}
for $k_{y}$.
Antinodal quasiparticles at ${\bf k}^{*}$ have
an energy very close $ W_0$, the maximum of
the DDW gap. Hence, they can scatter
into a nearby wavevector while
creating a particle-hole pair near
the Fermi arcs (the inner section
of the hole pockets). This is very
different from the situation in the
$d$-wave superconducting state, where
there are only Fermi points, not arcs,
as a result of which, there is very little
phase space for low-energy particle-hole pairs.
Secondly, the density of states is enhanced at the
gap edge, resulting in an
abundance of available phase space into which
the quasiparticle can be scattered. In
the superconducting state, this density of
states enhancement is cancelled by
coherence factors. These coherence
factors reflect the fact that the
quasiparticles are neutral, so they
are only weakly scattered by interactions
which are coupled to charge. 

We will set up the lifetime calculation in full generality, assuming
that both DDW and DSC order parameters are present, and then
vary the size of the DSC order parameter.
In order to more easily compare with experimental
results, we will assume mean-field-like
temperature dependence for the DSC gap so that
we can display our results as a temperature-dependent
decay rate.

Consider an initial quasiparticle state 
of momentum ${\bf k_1}$ in the antinodal region (to be precise, ${\bf k}^{*}$ defined above)
and of energy ${\mathcal E}_{\mathbf{k}_{1}}$, where ${\mathcal E}_{\mathbf{k}} = \sqrt{{\left({E_-}(\mathbf{k})-\mu\right)^2}+ |\Delta(\mathbf{k})|^2}$.
In the pseudogap state, where the $d$-wave superconducting order parameter $\Delta=0$,
${\mathcal E}_{\mathbf{k}}={E_-}(\mathbf{k})-\mu$.
Suppose that this initial state decays into a final state
of energy
${\mathcal E}_{\mathbf{k}_{2}}+{\mathcal E}_{\mathbf{k}_{3}}+{\mathcal E}_{\mathbf{k}_{4}}$.
In lowest order perturbation theory,
the decay rate for such a process is 
\begin{widetext}
\begin{equation}
\label{eqn:decay-rate-1}
\frac{1}{\tau_1} = 2\pi\int_{{\mathbf{k}_2}{\mathbf{k}_3}{\mathbf{k}_4}} {|M_{{\mathbf{k}_1}{\mathbf{k}_2}{\mathbf{k}_3}{\mathbf{k}_4}}|^2}
\:(2\pi)^{2}\delta({\bf k_1}-{\bf k_2}-{\bf k_3}-{\bf k_4})
\delta({\mathcal E}_{\mathbf{k}_{1}}-{\mathcal E}_{\mathbf{k}_{2}}-
{\mathcal E}_{\mathbf{k}_{3}}-{\mathcal E}_{\mathbf{k}_{4}})
[1-f({\mathcal E}_{\mathbf{k}_{2}})][1-f({\mathcal E}_{\mathbf{k}_{3}})]
[1-f({\mathcal E}_{\mathbf{k}_{4}})],
\end{equation}
\end{widetext}
where $\int_{{\mathbf{k}_2}{\mathbf{k}_3}{\mathbf{k}_4}} =\int\frac{{d^2}k_{2}}{(2\pi)^2}\frac{{d^2}k_{3}}{(2\pi)^2}
\frac{{d^2}k_{4}}{(2\pi)^2}$;  $M_{{\mathbf{k}_1}{\mathbf{k}_2}{\mathbf{k}_3}{\mathbf{k}_4}}$ is a matrix element, and $f({\mathcal E}_{\mathbf{k}})$is the Fermi function. We have in mind a situation in which ${\bf k_2}$
is close to ${\bf k_1}$ and ${\bf k_3},{\bf k_4}$
are close to the zone diagonal, but we will perform
the integrals over the full Brillouin zone. 

There is a second contribution to the decay rate, $1/\tau_{2}$,
resulting from scattering off thermally-excited quasiparticles.
The corresponding expression involves a different matrix element 
$N_{{\mathbf{k}_1}{\mathbf{k}_2}{\mathbf{k}_3}{\mathbf{k}_4}}$ and the quasiparticle at momentum ${\bf k_2}$
is thermally excited with probability
$f({\mathcal E}_{\mathbf{k}_{2}})$.
In all other respects the equation is the same as Eq.~\ref{eqn:decay-rate-1} except that the energy and momentum conserving  $\delta$-functions must be changed accordingly.The total decay rate is the sum $1/\tau=1/\tau_{1}+1/\tau_{2}$. 

The matrix elements
$M_{{\mathbf{k}_1}{\mathbf{k}_2}{\mathbf{k}_3}{\mathbf{k}_4}}$ and $N_{{\mathbf{k}_1}{\mathbf{k}_2}{\mathbf{k}_3}{\mathbf{k}_4}}$
will depend on the form
of the interaction between quasiparticles
and also on the coherence factors.
If we choose a Hubbard-like density-density interaction,
$\lambda \rho_{\uparrow}(\mathbf{q})\rho_{\downarrow}(-\mathbf{q})$, between 
the original electrons, the coherence factors
are extremely complicated in the coexisting DDW and DSC state,
and the multidimensional numerical integrations involved in calculating the scattering rates become next to impossible to carry out.
To obtain upper bounds, we replace them by their maximum values.  In the
DDW state, we expect the coherence factors
to be rather tame, but in the state with
both DSC and DDW orders, they will suppress
the decay rate as in a pure superconducting state.
Thus, such an approximation will 
{\it underestimate} the difference between
the decay rates in the pseudogap and the underdoped
superconducting states. We call this interaction, treated with
this approximation, model A.

In order to capture the effect of the coherence factors in the mixed DDW and DSC state, we
consider a model  interaction.  Since we are only concerned with the interaction between the quasiparticles in the valence band, we 
choose the interaction  to be ($\Omega$ is the volume of the system.)
\begin{eqnarray}
{\mathcal V}= \frac{\lambda}{\Omega} \sum_{\mathbf{q},\mathbf{k},\mathbf{k}'}\psi_{\mathbf{k}\uparrow}^{v\dagger} \psi_{\mathbf{k}+\mathbf{q}\uparrow}^{v} \psi_{\mathbf{k}'\downarrow}^{v\dagger} \psi_{\mathbf{k}'-\mathbf{q}\downarrow}^{v} ,
\end{eqnarray}
where $\psi_{\mathbf{k},\alpha}^{v\dagger}$ creates a
quasiparticle in the valence band of the DDW state. We
have ignored the temperature and momentum dependence of the interaction,
because we are primarily interested
in temperatures much lower than the DDW transition temperature, where
the temperature dependence of the DDW gap should be
weak. Moreover, we merely wish to demonstrate how the development
of superconductivity affects the lifetime, so we also neglect
the momentum space structure of the interaction. 
For this interaction, which we call model B, the coherence factors
are equal to unity for the DDW order
alone, but are non-trivial in the state with
both orders as a result of the coherence factors
associated with superconductivity in the mixed state. 
We can view the mixed state as DSC developing on top of DDW. Thus, the coherence factors for this interaction can be read off from the BCS theory and the matrix elements are
\begin{multline}
M_{{\mathbf k}_1,{\mathbf k}_2,{\mathbf k}_3,{\mathbf k}_4} =\lambda [-V_{{\mathbf k}_1}V_{{\mathbf k}_2}U_{{\mathbf k}_3}V_{{\mathbf k}_4}
+ V_{{\mathbf k}_1}U_{{\mathbf k}_2}V_{{\mathbf k}_3}V_{{\mathbf k}_4}\\
 - U_{{\mathbf k}_1}V_{{\mathbf k}_2}U_{{\mathbf k}_3}U_{{\mathbf k}_4}
+ U_{{\mathbf k}_1}V_{{\mathbf k}_2}V_{{\mathbf k}_3}U_{{\mathbf k}_4}]
\end{multline}
and
\begin{equation}
\begin{split}
N_{{\mathbf k}_1,{\mathbf k}_2,{\mathbf k}_3,{\mathbf k}_4}=&\lambda[-V_{{\mathbf k}_1}V_{{\mathbf k}_2}V_{{\mathbf k}_3}V_{{\mathbf k}_4}-U_{{\mathbf k}_1}U_{{\mathbf k}_2}U_{{\mathbf k}_3}U_{{\mathbf k}_4}\\
&-U_{{\mathbf k}_1}V_{{\mathbf k}_2}V_{{\mathbf k}_3}U_{{\mathbf k}_4}-V_{{\mathbf k}_1}U_{{\mathbf k}_2}U_{{\mathbf k}_3}V_{{\mathbf k}_4}\\
&+U_{{\mathbf k}_1}V_{{\mathbf k}_2}U_{{\mathbf k}_3}V_{{\mathbf k}_4}
-V_{{\mathbf k}_1}U_{{\mathbf k}_2}U_{{\mathbf k}_3}V_{{\mathbf k}_4}\\
&-U_{{\mathbf k}_1}V_{{\mathbf k}_2}V_{{\mathbf k}_3}U_{{\mathbf k}_4}
+V_{{\mathbf k}_1}U_{{\mathbf k}_2}V_{{\mathbf k}_3}U_{{\mathbf k}_4}]
\end{split}
\end{equation}
where
\begin{equation}
\begin{array}{c}
U_{\mathbf{k}}^{2}\\
V_{\mathbf{k}}^{2}
\end{array}\bigg\}=\frac{1}{2}\,\left(1\pm\frac{{E_-}(\mathbf k)-\mu}{{\mathcal E}(\mathbf{k})}\right).
\end{equation}
This form of the interaction allows us capture the difference between
the matrix elements in the pseudogap and superconducting
states.
\begin{figure}[htb]
\includegraphics[scale=0.3]{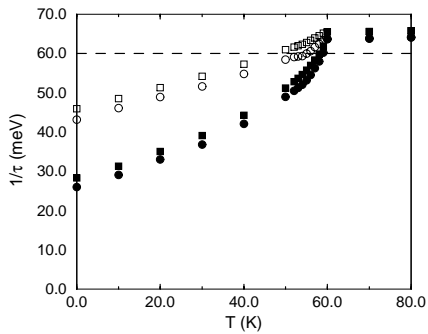}
\caption{The lifetime of a quasipartcle in the antinodal region at wave vector ${\mathbf k}^{*}$, as defined in the text, plotted as a function of temperature. The open symblos are for without coherence factors and the solid symbols are with coherence factors. The square symbols correspond  to 2 meV broadening of the energy conserving $\delta$-function and the circles to 1 meV broadening. The electron-electron interaction parameter $\lambda = 1.5$ eV. }
\label{fig:band1}
\end{figure}

Our results are displayed in Fig.~ \ref{fig:band1}, where
the decay rates are plotted against temperature.
In these plots we have kept the total gap ${\mathcal E}_{k^{*}}$ fixed and equal to $0.06$ eV, while assuming a  mean-field temperature dependence for
the superconducting gap 
\begin{equation}
\Delta_{0}(T) =
\Delta_{0}(0)\left(1-\frac{T}{T_c}\right)^{1/2}.
\end{equation}
with $\Delta_{0}(0)=0.03$ eV and $T_{c}=60$ K. This implicitly defines the temperature dependence of the DDW gap, which is weak close to $T_{c}$, as noted earlier.

It is apparent from this figure that the decay rate drops
dramatically as a result of the development of
superconducting order. From the calculation for
model A, we see that there is a substantial drop
resulting from the elimination of phase space.
From model B, we see that the coherence factors
reduce the decay rate further by a large amount.

The absence of an antinodal quasiparticle
peak in the pseudogap state and its subsequent
emergence in the superconducting state has
been interpreted here as the increase of its width
 as $T_c$ is approached.
However,  when the width
becomes comparable to the quasiparticle energy,
i.e. as the curve reaches the dashed line in Figs.~ \ref{fig:band1},
the quasiparticle concept breaks down. Once this occurs,
our perturbative calculation can no longer be trusted,
and it is not meaningful to assign a width or
weight to the quasiparticle. What is significant is that it {\it is}
possible
to have a reasonably well-defined quasiparticle in the
superconducting state as a result of
phase space restrictions and coherence factors,
as we have found. 

The  broken symmetry state
may or may not have well-defined quasiparticles at
the single-particle gap edge. It depends on many non-universal details: the locus in momentum space,  the doping level, the  interaction strength,
etc.  Thus, the absence of a well defined
antinodal quasiparticle does not preclude
a broken symmetry state. However, it may have important
effects on non-universal aspects of the state, such as
the temperature dependence of the order parameter
which may, as a result, be strongly non-mean-field-like.
Also, our calculation leaves out
 fluctuation effects, which
must be considered in the future.

We end with three concluding remarks: (1) Although hole pockets cannot be observed in ARPES, other experimental probes can be used to look for their signature, for example,  infrared Hall angle measurement in the underdoped regime.\cite{Dennis}  (2) Because the interlayer tunneling matrix element is so strongly peaked at $(\pi,0)$,\cite{Chakravarty} we expect the $c$-axis optical conductivity to show a strong temperature dependence at $T_{c}$ given our lifetime calculation. Indeed, this is consistent with the known measurements.\cite{Basov} (3) We have not yet studied in detail the doping dependence. Nonetheless, it is possible to make a qualitative observation. There are two competing effects. As the system is moderately underdoped, the DDW order parameter must increase. Thus, the quasiparticle in the $(\pi, 0)$ regime will have higher energy, increasing its scattering rate. On the other hand, the Fermi arcs will also shrink and the phase space for particle-hole scattering will decrease.  We wish to return to these issues in the near future.

C. N. has been supported by the NSF under
Grant No. DMR-9983544 and the A.P. Sloan Foundation.
S. C. and S. T. have been supported by the NSF under
Grant No. DMR-9971138. We would like to thank Peter Armitage, Dimitri Basov, John F{\ae}restad, Eduardo Fradkin, Koichi Hamada, Jiangping Hu, Steven Kivelson, Richard Thompson, Douglas Scalapino, Daijiro Yoshioka, and  Jan Zaanen for discussions.

\end{document}